\documentclass[a4paper,orivec,runningheads]{llncs}

\usepackage{ifpdf}

\usepackage{amsmath} 
\usepackage{amssymb}

\usepackage[utf8]{inputenc}

\usepackage[T1]{fontenc}

\usepackage{textcomp}

\usepackage{graphicx}

\usepackage{multirow}

\usepackage{diagbox}

\DeclareMathOperator{\Id}{Id}

\DeclareMathOperator{\diag}{diag}

\providecommand{\norm}[1]{\left\lVert#1\right\rVert}

\usepackage{xspace}

\makeatletter
\DeclareRobustCommand\onedot{\futurelet\@let@token\@onedot}
\def\@onedot{\ifx\@let@token.\else.\null\fi\xspace}

\providecommand{\ie}{i.e\onedot}

\makeatother

\providecommand{\Figref}[1]{Figure~\ref{#1}}

\usepackage{amssymb}
\usepackage{makeidx}  
\usepackage{url}     
\usepackage{amsmath}
\usepackage{nccmath}
\usepackage{amsfonts} 
\usepackage{multirow}
\usepackage{mathrsfs}
\usepackage{siunitx}
\usepackage{placeins}
\usepackage{esvect}
\usepackage{pbox}

\usepackage{color,soul}
\DeclareFontFamily{OT1}{pzc}{}
\DeclareFontShape{OT1}{pzc}{m}{it}{<-> s * [1.10] pzcmi7t}{}
\DeclareMathAlphabet{\mathpzc}{OT1}{pzc}{m}{it}

\newcommand{\myparagraph}[1]{\paragraph{\textbf{#1.}}}
\DeclareMathOperator{\FWHM}{FWHM}

\graphicspath{{pics/}{figs/}}

\usepackage{tikz}
\usetikzlibrary{arrows}
\usetikzlibrary{calc}
\usetikzlibrary{shapes.misc}
\pgfdeclarelayer{background}
\pgfdeclarelayer{foreground}
\pgfsetlayers{background,main,foreground}

\usepackage{hyperref}

\begin{document}
\frontmatter          
\mainmatter              
\title{Scale factor point spread function matching: beyond aliasing in image resampling}%
\titlerunning{Scale factor PSF matching: beyond aliasing in image resampling}%
\author{M. Jorge Cardoso\inst{1}${}^{,}$\inst{2}\and Marc Modat\inst{1}${}^{,}$\inst{2}\and Tom Vercauteren\inst{1}\and \\Sebastien Ourselin\inst{1}${}^{,}$\inst{2} }
\index{Cardoso, M. Jorge} 
\index{Modat, Marc} 
\index{Vercauteren, Tom} 
\index{Ourselin, Sebastien}
\authorrunning{M.J. Cardoso \textit{et al.}}

\institute{Translational Imaging Group, CMIC, University College London, UK \and Dementia Research Centre (DRC), University College London, UK}

\maketitle

%
\begin{abstract}%
Imaging devices exploit the Nyquist-Shannon sampling theorem to avoid both aliasing and redundant oversampling by design.
Conversely, in medical image resampling, images are considered as continuous functions, are warped by a spatial transformation, and are then sampled on a regular grid.
In most cases, the spatial warping changes the frequency characteristics of the continuous function and no special care is taken to ensure that the resampling grid respects the conditions of the sampling theorem. This paper shows that this oversight introduces artefacts, including aliasing, that can lead to important bias in clinical applications.
One notable exception to this common practice is when multi-resolution pyramids are constructed, with low-pass "anti-aliasing" filters being applied prior to downsampling.
In this work, we illustrate why similar caution is needed when resampling images under general spatial transformations and
propose a novel method that is more respectful of the sampling theorem, minimising aliasing and loss of information.
We introduce the notion of scale factor point spread function (sfPSF) and employ Gaussian kernels to
achieve a computationally tractable resampling scheme that can cope with arbitrary non-linear
spatial transformations and grid sizes. Experiments demonstrate significant ($p<10^{-4}$) technical and clinical implications of the proposed method.
\end{abstract}
%
%
%
\section{Introduction}
Image resampling is ubiquitous in medical imaging. Any 
processing pipeline that requires coordinate mapping, 
correction for imaging distortions, or simply altering the 
resolution of an image, needs a representation of the image 
outside of the digital sampling grid provided by the initial 
image.
The theoretical foundations from the Nyquist-Shannon 
sampling theorem provide conditions under which perfect 
continuous reconstruction can be achieved from regularly 
spaced samples.
This theoretical model thus provides a means of interpolating 
between the discrete samples and underpins the typical 
resampling procedure used in medical imaging.
%
%
Many interpolating methods, ranging from low order, e.g. piecewise constant and linear interpolation, to high order, e.g. polynomial, piecewise-polynomial (spline) and windowed sinc (Lanczos) methods have thus been developed
, with different methods being optimal for specific types of signal.

Resampling typically relies on one such interpolation method to represent, on a given (discrete) target sampling grid, a continuous image derived from the input source image. In most cases, the (discrete) resampled image is simply assumed to be a faithful representation of the continuous derived image, with no special care taken to ensure the sampling theorem conditions.
Within the context of medical imaging, Meijering \textit{et al.} \cite{Meijering:1999gi} compared multiple resampling kernels, concluding that the quality of the resampling was proportional to the degree of the kernel, with linear, cubic and Lanczos resampling kernels being the best with a $1^{st}$, $2^{nd}$ and $3^{rd}$ order neighbourhood respectively.
This comparison was done with grid aligned data under translation, thus respecting the Nyquist criterion, meaning that their conclusions cannot be extrapolated to general resampling.

Resampling is known to introduce artefacts in medical imaging.
One common example arises within the context of image registration \cite{Pluim:1999gx,Aganj:2013ho} where a source image is typically resampled to the sampling grid of a target image and a similarity function is used to compare the images based on the discrete samples at the target grid location only, thereby discarding the continuous image representation.
%
For example, Pluim \textit{et al.} \cite{Pluim:1999gx} ameliorated the problem by using partial volume (PV) sampling, and more recently Aganj \textit{et al.} \cite{Aganj:2013ho} replaced the cost function summation term with an approximate integral.
Another main source of artefacts is the fact that aliasing is introduced when resampling an image to lower resolution.
While, in multi-resolution pyramids \cite{Lindeberg:PAMI:1990}, aliasing is well addressed by applying a Gaussian filter prior to downsampling, the problem in the general resampling case has surprisingly received little attention in the medical imaging community.
Unser \textit{et al.} \cite{Unser:ICIP:1994} propose a least-squares spline-based formulation that achieves aliasing-free resampling for affine transformations, but to the best of our knowledge no general solution has been proposed for local transformations.
Inverse approximation-based resampling has also been proposed ~\cite{Arigovindan:TIP:2005}, but this technique cannot easily take into account the local anisotropy of the transformations and has the drawback of requiring the inverse transformation. 

In this paper, we propose to address aliasing artefacts in the general context of resampling with non-linear spatial transformations by associating a scale factor point spread function (sfPSF) to the images and formulating the resampling process as an sfPSF matching problem.
In the case of an original image from a known imaging device, the sfPSF is simply the PSF of the device.
%
Our approach models local sfPSF transformations as a result of coordinate mapping, providing a unified and scale-aware way to resample images between grids of arbitrary sizes, under arbitrary spatial transformations.

%

%
%
%

%
\section{Methods and Methodological considerations}
\label{MethodCons}

\myparagraph{Nominal scale factor Point Spread Function (sfPSF)}
Observed samples (pixels/voxels) in medical images are not direct Dirac impulse measurements of a biological scene, but are typically samples from the convolution of the unobservable biological scene with an observation kernel, i.e. they have an associated point spread function (PSF).
This PSF is commonly a property of the acquisition or reconstruction system. For example, 3D MRI images have approximately Gaussian PSFs due to several factors, including the low-pass filters in the receiver coils, while CT and PET images have spatially variant PSFs dependent on many parameters including the image reconstruction algorithm.

In this work, we propose to associate a PSF to each image, whether it directly arises from an imaging device or is the result of further transformation or processing, and refer to this as the scale factor PSF (sfPSF) by analogy with the scale factor used in scale space theory \cite{Lindeberg:PAMI:1990}.
In the case where the image directly arises from an imaging device with known PSF, the sfPSF is simply taken as the closest Gaussian approximation.
When the actual PSF associated with a given image is unkown, without loss of generality the nominal sfPSF is assumed to be a Gaussian $\mathcal{G}_\Sigma$, where the covariance matrix $\Sigma$ is diagonal and the full-width-half-maximum (FWHM) of the Gaussian PSF is matched to the voxel size $D=\{D_x,D_y,D_z\}$:
$\Sigma=(2\sqrt{2\log2})^{-2}\diag(D_x^2,D_y^2,D_z^2)$.
The relevance of the sfPSF will become clear in the following section. Intuitively, when an image with an associated sfPSF is smoothed, the resulting smoothed image will have larger sfPSF. Conversely, when an image is upsampled, the resulting image will be associated with the same sfPSF as the source image (in world coordinates).
%
\myparagraph{Compensating for aliasing when downsampling}
In medical images, frequencies commonly approach the Nyquist limit of their representation (e.g. full k-space sampling in MRI).
If one affinely transforms an image onto its own grid with an affine transformation which has a determinant greater than 1, i.e. spatially compressing the samples, or transforms an image into a lower resolution grid, then the Nyquist criterion is not satisfied with typical interpolation. This sub-Nyquist sampling introduces frequency aliasing and loss of information. 
\begin{figure}[b!]
	\vskip -10pt
	\tiny
	\begin{center}
	\begin{tabular}{cccccccc}
	High Res&\multirow{2}{*}{$\circledast$}&Resampling&\multirow{2}{*}{$\rightarrow$}&Resampled to&\multicolumn{3}{c}{Resampled to Low Resolution Grid}\\
	Image&&Grid&&High Res Grid&Linear&Sinc&Gaussian sfPSF\\
\includegraphics[interpolate=false,width=0.13\textwidth]{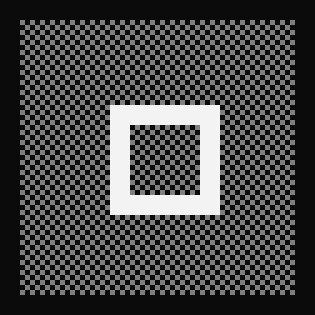}&$\circledast$&
\includegraphics[interpolate=false,width=0.13\textwidth]{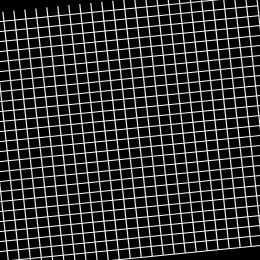}&$\rightarrow$&
\includegraphics[interpolate=false,width=0.13\textwidth]{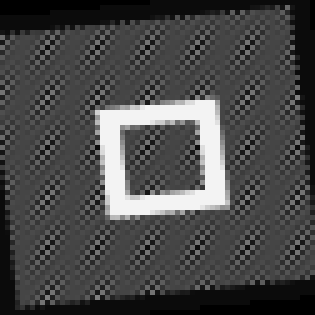}&
\includegraphics[interpolate=false,width=0.13\textwidth]{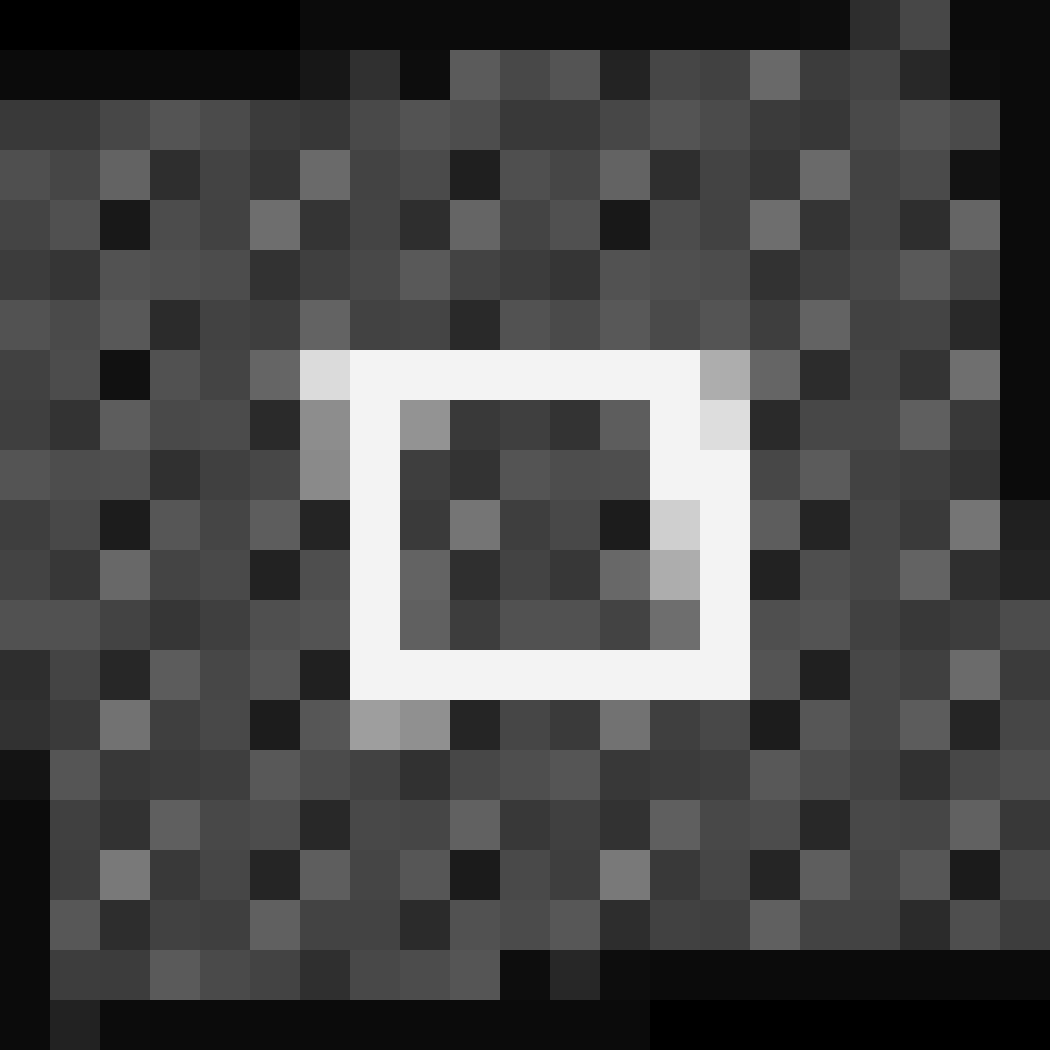}&
\includegraphics[interpolate=false,width=0.13\textwidth]{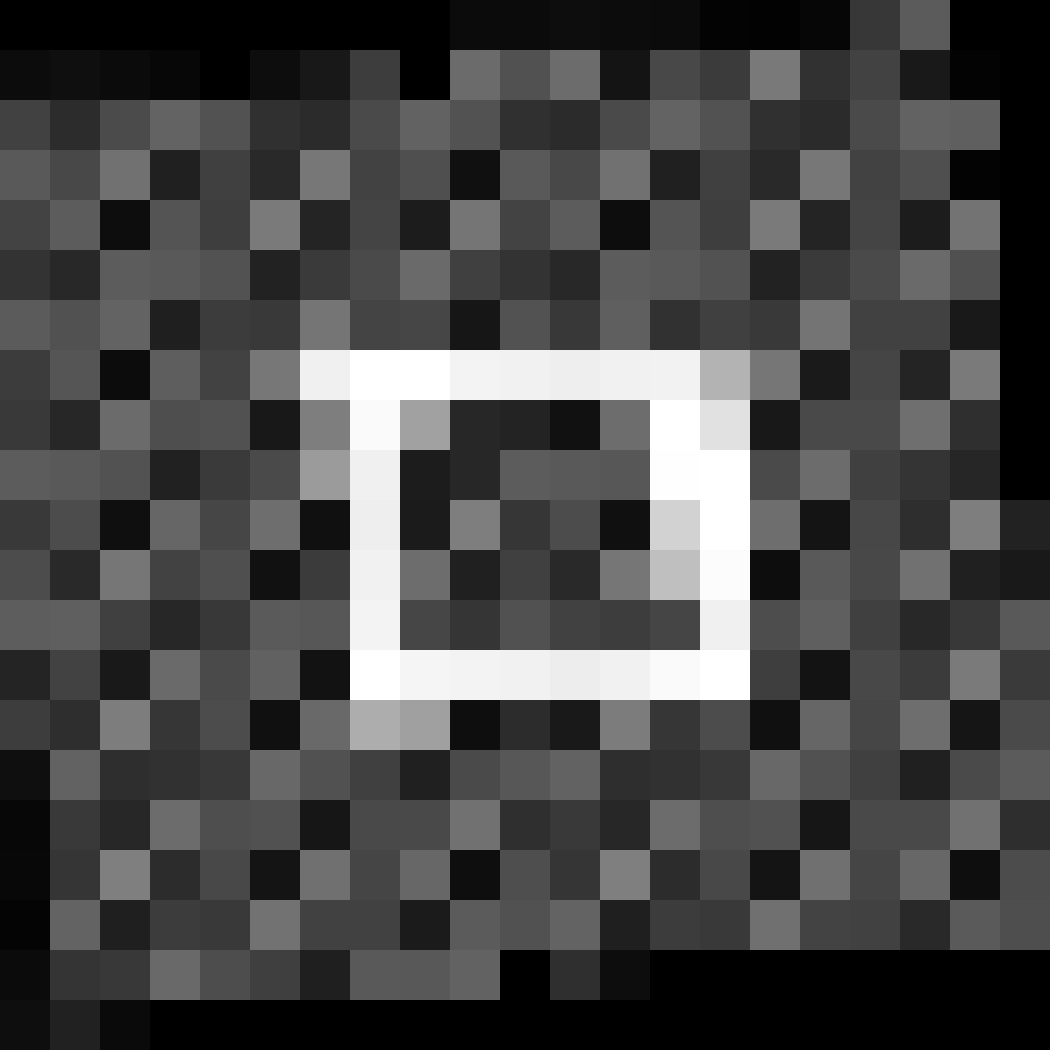}&
\includegraphics[interpolate=false,width=0.13\textwidth]{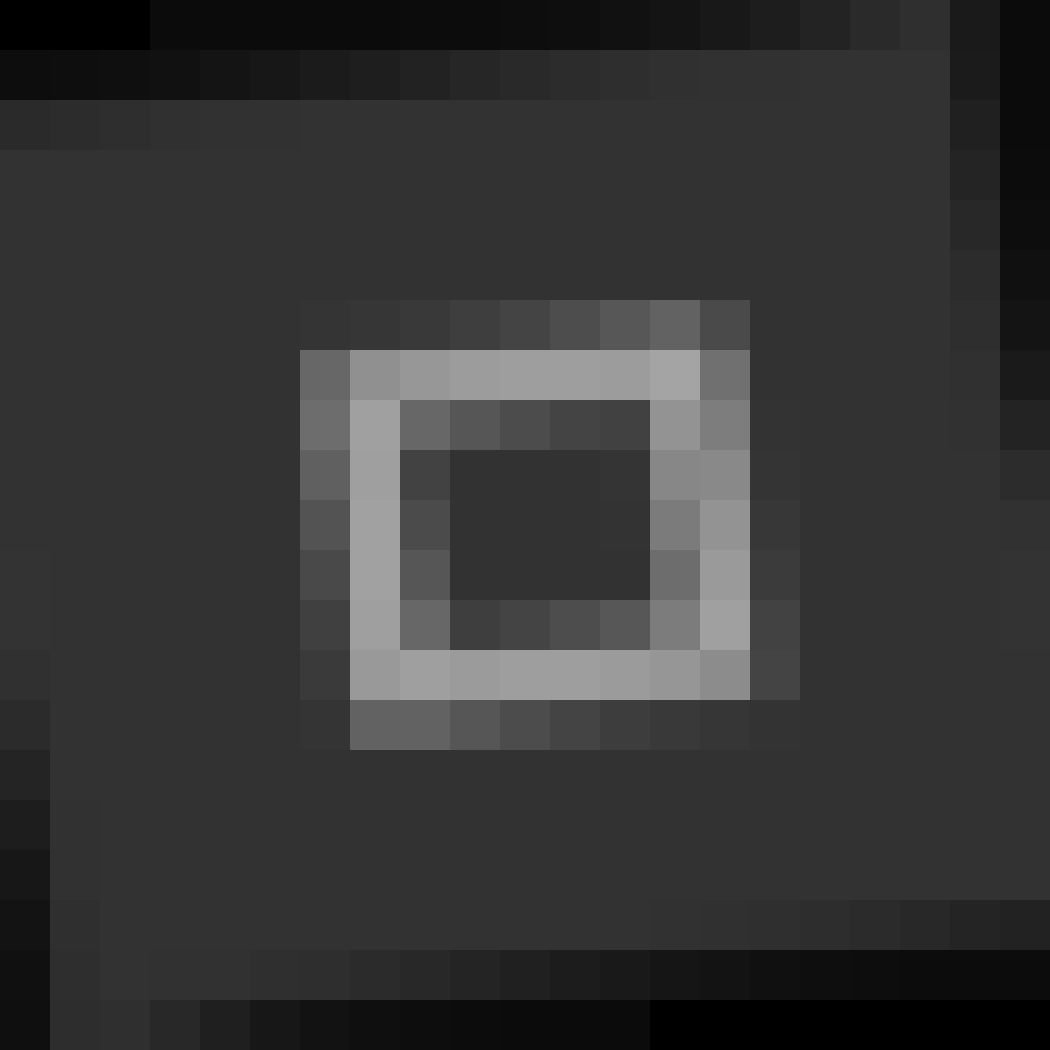}\\
\includegraphics[interpolate=false,width=0.13\textwidth]{Code/test_hr.png}&&
\includegraphics[interpolate=false,width=0.13\textwidth]{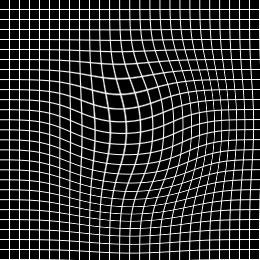}&&
\includegraphics[interpolate=false,width=0.13\textwidth]{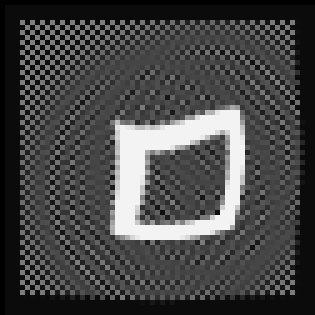}&
\includegraphics[interpolate=false,width=0.13\textwidth]{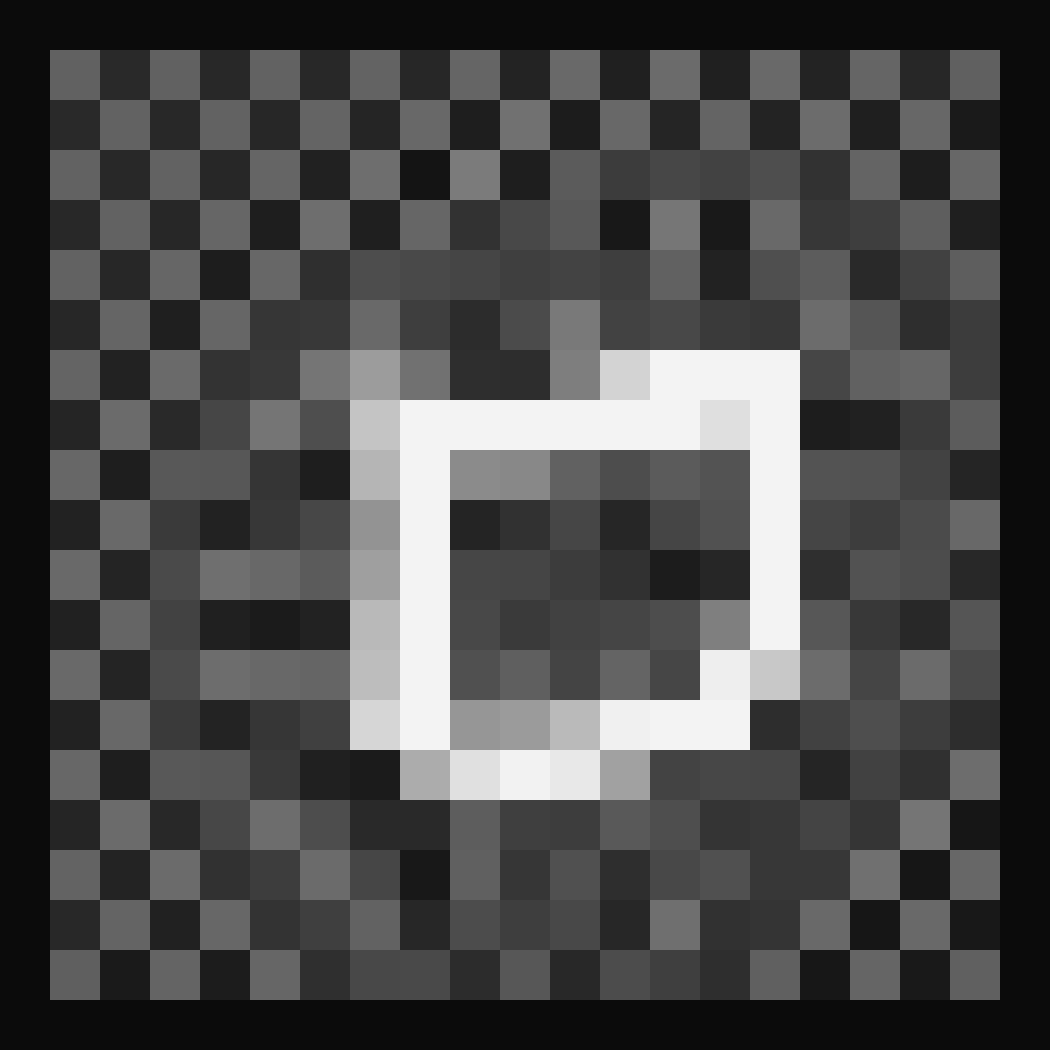}&
\includegraphics[interpolate=false,width=0.13\textwidth]{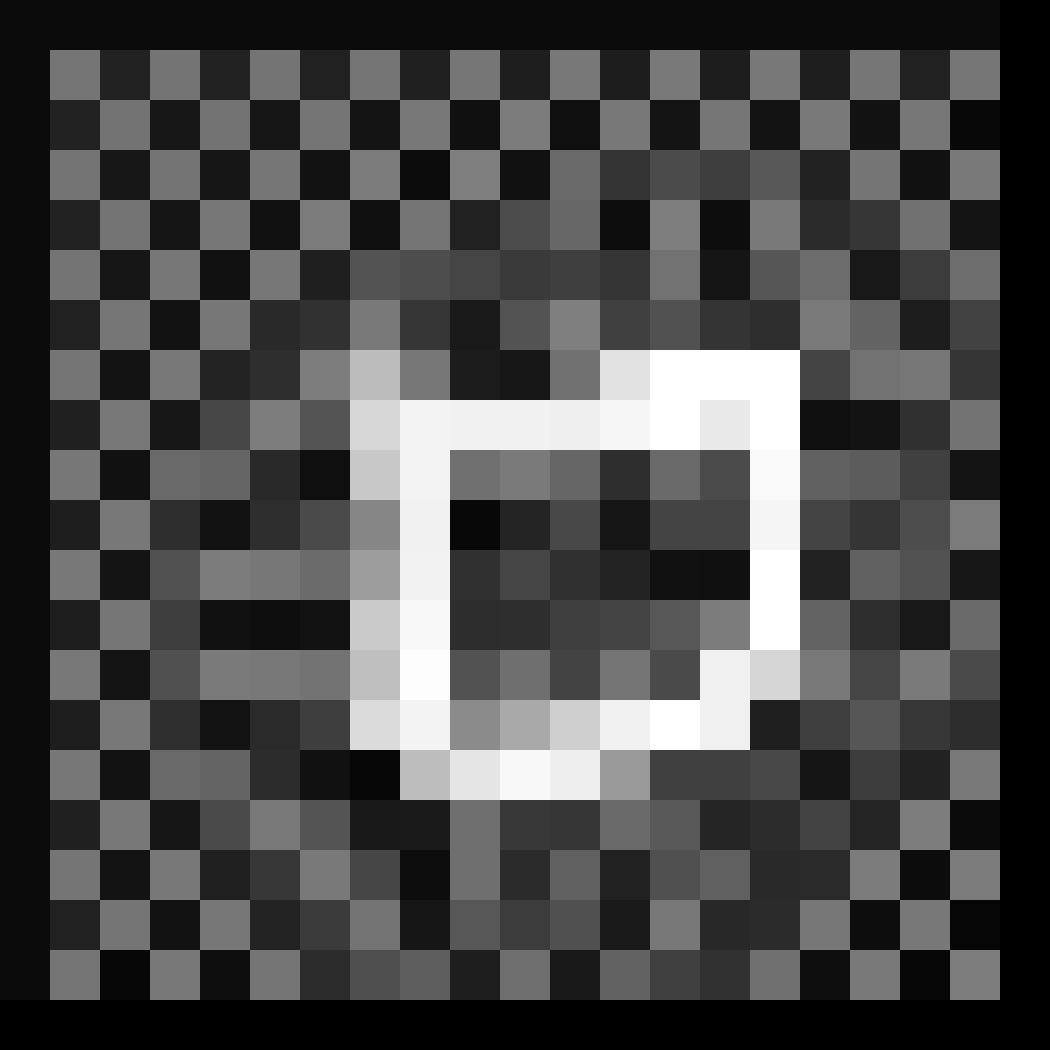}&
\includegraphics[interpolate=false,width=0.13\textwidth]{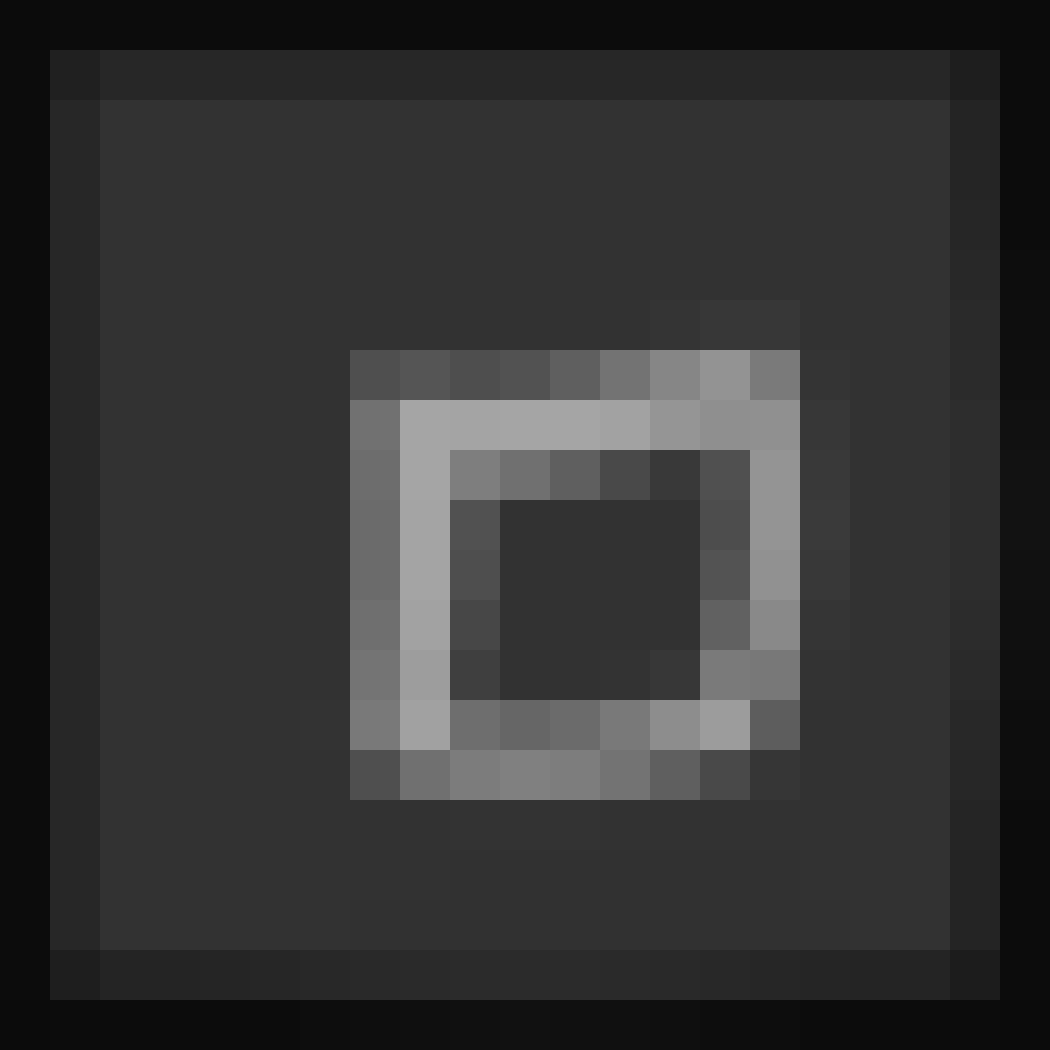}
\end{tabular}
	\end{center}
	\normalsize
	\vskip -10pt
    \caption{Left to right: a high resolution (HR) image (63x63 voxels at 1x1$\textit{mm}$ voxel size) with a square on a high frequency pattern; the resampling grid (transformation); the HR image resampled to the original HR grid and to a lower resolution (21x21 voxels at 3x3$\textit{mm}$ voxel size) grid using linear, sinc and a Gaussian sfPSF (proposed) interpolation. Note the aliasing in the linear/sinc examples.}
	\vskip -15pt
	\label{Toy_Example}
\end{figure}

In scale-space and pyramidal approaches, when downsampling an image by a factor of $k=2$, the original image is typically pre-convolved with a Gaussian filter with variance $\sigma_P=0.7355$.
%
Applying the notion of sfPSF and the notations introduced above, we specify
$\mathcal{G}_{{\Sigma}_S}$ and $\mathcal{G}_{{\Sigma}_T}$ as the source image and target grid nominal sfPSFs respectively.
Following the scale space methodology \cite{Lindeberg:PAMI:1990}, sfPSF matching consists of finding the best scale-matching PSF $\mathcal{G}_{{\Sigma}_P}$ that, when combined with the source sfPSF, provides the best approximation of the target sfPSF.
%
%
If the sfPSFs are Gaussian, then the covariance of $\mathcal{G}_{{\Sigma}_P}$ can be obtained through simple covariance addition ${{{\Sigma}_T}={{\Sigma}_S}}+{{\Sigma}_P}$.
In the downsampling example above, without additional knowledge, we have $\FWHM_T=k=2$ and $\FWHM_S=1$.
We therefore get $\Sigma_P=\Sigma_T-\Sigma_S=(k^2-1^2)(2\sqrt{2\log{2}})^{-2}\Id$ or, in this case, $\sigma_P=0.7355$.
%
This is equivalent to finding the best Fourier domain filter that limits the bandwidth of the source image frequencies to the representable Nyquist frequencies on the target grid.
Similar filtering would be necessary to avoid aliasing when rigidly resampling high resolution images to a lower resolution, e.g. high resolution anatomical MRI and CT images or binary segmentations are transformed to lower resolution MRI (2D FLAIR), metabolic (PET), microstructural (DWI) and functional (fMRI) images.
In this work we propose to compensate for the differences in sfPSF when resampling to arbitrary grid sizes under global (affine) or local (non-linear) spatial transformations. 

Before presenting our methodological contribution in detail,
we illustrate the sfPSF performance in comparison
with standard interpolation on
a synthetic phantom (Fig.~\ref{Toy_Example}) with a low frequency object (square) overlaid on a high-frequency pattern (to highlight aliasing).
This phantom is resampled to a grid with 3 times less resolution \textit{per} axis. Note that the proposed sfPSF method appropriately integrates out high frequencies without introducing aliasing. 

%
\myparagraph{Pre-convolution sfPSF matching}
In the affine transformation $\mathcal{A}$ scenario from the space of $T$ to $S$,
the source image can be pre-convolved with an
\emph{anti-aliasing} filter with covariance $\Sigma_P^S+\Sigma_S=\mathcal{A}\cdot\Sigma_T\cdot\mathcal{A}^{\intercal}$ and then resampled using classical interpolation. Note that  $\Sigma_{P}^S$ is now defined in the space of $S$.

In the general non-linear spatial transformation case, a similar but spatially-variant pre-filtering procedure could be applied by relying on a local affine approximation of the spatial transformation, i.e. using the jacobian matrix.
This solution might provide useful results but because the sfPSF matching, or smoothing, is performed before the spatial transformation, and the sfPSF potentially has a large spatial extent, it can be seen as disregarding the non-homogeneous distribution of the samples in target space $T$.
%
Instead, we propose to model and perform the sfPSF matching in the space of $T$ and thus rely on $\Sigma_{P}^T$. 

%
\myparagraph{Target space sfPSF matching}
%
%
Let $\mathcal{F}_{S\leftarrow T}(\mathbf{v})$ be a spatial transformation mapping the location $\mathbf{v}=(v_x,v_y,v_z)$ from the space of $T$ to that of $S$, given either an affine or non-linear transformation. 
Let $\mathcal{A}_{S\leftarrow T}(\mathbf{v})$ be the jacobian matrix of $\mathcal{F}$ at $\mathbf{v}$ which provides the best linear approximation of $\mathcal{F}$ at $\mathbf{v}$.
Following our previous derivations we have $\Sigma_P^S(\mathbf{v})+\Sigma_S \approx \mathcal{A}_{S\leftarrow T}(\mathbf{v})\cdot\Sigma_T\cdot\mathcal{A}_{S\leftarrow T}^\intercal(\mathbf{v})$.
Thus, given $\Sigma_P^S(\mathbf{v})=\mathcal{A}_{S\leftarrow T}(\mathbf{v})\cdot\Sigma_P^T(\mathbf{v})\cdot\mathcal{A}_{S\leftarrow T}^\intercal(\mathbf{v})$
we now want
$\Sigma_P^T(\mathbf{v})+\mathcal{A}_{S\leftarrow T}^{-1}(\mathbf{v})\cdot\Sigma_S\cdot\mathcal{A}_{S\leftarrow T}^{-\intercal}(\mathbf{v}) \approx \Sigma_T$.
In other words, we want to find a \emph{symmetric positive semi-definite} covariance $\Sigma_P^T(\mathbf{v})$ that, when combined with the affinely transformed covariance
$\Sigma_S^T(\mathbf{v}) = \mathcal{A}_{S\leftarrow T}^{-1}(\mathbf{v})\cdot\Sigma_S\cdot\mathcal{A}_{S\leftarrow T}^{-\intercal}(\mathbf{v})$, best approximates $\Sigma_T$. 
%
%
Under the typical assumption that the nominal sfPSF $\Sigma_T$ is diagonal, this approximation is given by $\tilde{\Sigma_P^T}(\mathbf{v})=\max(\Sigma_T-\lambda(\mathbf{v}), 0)$ with $\lambda(\mathbf{v})$ being a diagonal scaling matrix containing the components of $\Sigma_S^T(\mathbf{v})$ obtained through polar decomposition
and $max(\cdot,\cdot)$ is the element-wise maximum operator between two matrices. It is important to note that under regimes where the Nyquist limit is not violated, e.g. upsampling and rigid transformation between isotropic grids, the proposed method reverts to standard resampling as the sfPSF would become a Dirac.

%
\myparagraph{Interpolation by convolution with the sfPSF}
%
%
Let $I_T(\mathbf{v})$ be the sought resampled intensity in the space of $T$ at location $\mathbf{v}$, and $I_S(\mathbf{v})$ be the intensity of the source image.
%
%
The interpolated and PSF matched value of $I_T$ at location $(\mathbf{v})$ can be obtained by an \emph{oversampled} discretised convolution 
\vskip -5pt
\begin{align}
I_T(\mathbf{v})&=\frac{1}{Z}\sum_{\mathpzc{v}_x}^{N_x(\mathbf{v})}\sum^{N_y(\mathbf{v})}_{\mathpzc{v}_y} \sum^{N_z(\mathbf{v})}_{\mathpzc{v}_z} I_S(\mathcal{F}_{S\leftarrow T}(\mathbf{v}-\mathbf{\mathpzc{v}}))\mathcal{G}_{\tilde{\Sigma^T_P}{(\mathbf{v})}}(\mathbf{\mathpzc{v}})
\label{convRes}
\end{align}
\vskip -5pt
\noindent where $\mathcal{G}_{\tilde{\Sigma^T_P}{(\mathbf{v})}}(\mathbf{\mathpzc{v}})=(2\pi)^{-\frac{3}{2}}|\tilde{\Sigma_P}{(\mathbf{v})}|^{-\frac{1}{2}}\, e^{ -\frac{1}{2}\mathbf{\mathpzc{v}}'(\tilde{\Sigma^T_P}{(\mathbf{v})})^{-1}\mathbf{\mathpzc{v}} }$, $Z$ is a normaliser that ensures the (discrete) sum over $\mathcal{G}_{\tilde{\Sigma^T_P}{(\mathbf{v})}}(\mathbf{\mathpzc{v}})$ equals 1.
$N_x(\mathbf{v})$, $N_y(\mathbf{v})$ and $N_z(\mathbf{v})$ are sampling regions centered at $\mathbf{v}$ in the $x$, $y$ and $z$ directions respectively. 
More specifically, $N_x(\mathbf{v})$ are homogeneously spaced samples between -3 and 3 standard deviations, i.e. $N_x(\mathbf{v})=\left\{k\sigma_{P_x}(\mathbf{v}) \ \ \forall k \in [-3,3] \right\}$,
with $\sigma_{P_x}^2(\mathbf{v})=\tilde{\Sigma^T_P}^{xx}(\mathbf{v})$,
and equivalently for $N_y(\mathbf{v})$, $N_z(\mathbf{v})$, $\sigma_{P_y}(\mathbf{v})$  and $\sigma_{P_z}(\mathbf{v})$. 
%
%
As S is a discrete image, an appropriate interpolation method, such as linear, cubic or sinc interpolation is used to get the sample values at $\mathcal{F}_{S\leftarrow T}(\mathbf{v}-\mathbf{\mathpzc{v}})$. Similarly, interpolation is required to compute $\mathcal{F}_{S\leftarrow T}(\mathbf{v}-\mathbf{\mathpzc{v}})$ if the spatial transformation is represented as a displacement field.
Note also that when $\sigma_{P_x}$ (resp. $\sigma_{P_y}$ or $\sigma_{P_z}$) becomes $0$ or very small, care has to be taken to appropriately compute the limit of the Gaussian weight by resorting to an axis aligned Dirac function.

\myparagraph{Limitations of Sinc interpolation by convolution}
Under the sampling theorem, replacing the Gaussian in Eq.\ref{convRes} with Sinc should produce the theoretically best alias-free results. Knowing that this work focuses on obtaining realistic PSF’s rather than the least aliased one, we note 3 problems with Sinc resampling that are addressed with the proposed method: First, Sinc is only optimal if the discretised sampling scheme of Eq.\ref{convRes} is replaced with an integral. Second, sincs are optimal from a bandlimit point of view, but are poor approximations of real PSFs, e.g. PET images have aproximate Gaussian PSF larger than the voxel size. Finally, sincs cannot be used for signal-limited images (probabilistic segmentations bounded from 0 to 1), as it would produce negative probabilities. Conversely, Gaussians do not have this problem.

%
\section{Validation and Discussion}
\myparagraph{Data} 30 healthy control subjects were obtained from the ADNI2 database. All subjects had an associated T1-weighted MRI image, acquired at 1.1$\times$1$\times$1$mm$ voxel size, and a ${}^{18}$F-FDG PET image, reconstructed at 3$\times$3$\times$3$mm$. For this selected subset of ADNI2, all data (MRI and PET) were acquired on the same scanner with the same scanning parameters, thus removing acquisition confounds. While the effective PSF of PET images can be between 3 and 6$mm$ FWHM, in these experiments we will assume a PSF with 3$mm$ FWHM. 

\begin{figure}[b!]
	\vskip -10pt
	\begin{center}
	\begin{tabular}{ccccc}
	\includegraphics[trim={0cm 0cm 0cm 1.1cm},clip,width=0.16\textwidth]{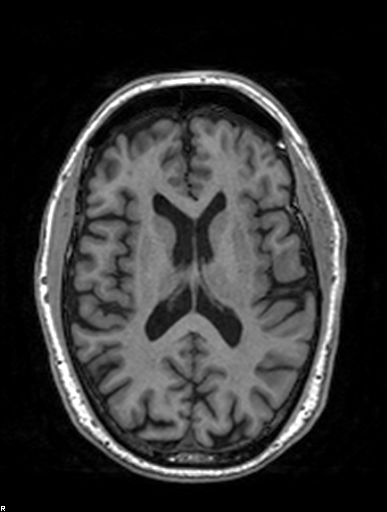}&
	\includegraphics[trim={1cm 1cm 1cm 2cm}, clip, width=0.18\textwidth]{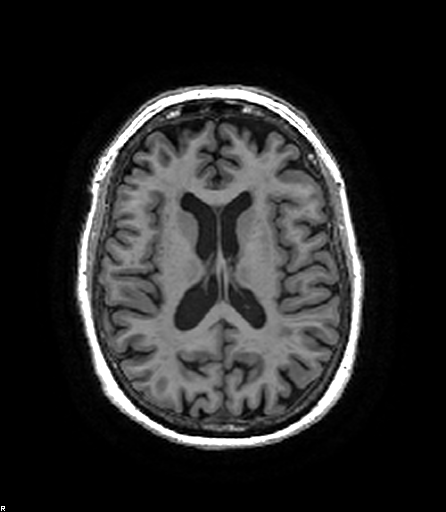}&
	\includegraphics[trim={1cm 1cm 1cm 2cm}, clip,width=0.18\textwidth]{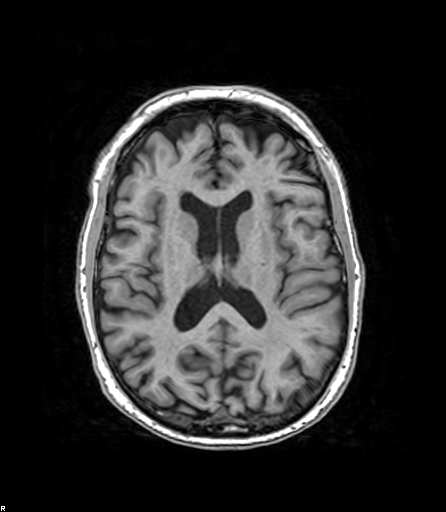}&
	\includegraphics[trim={1cm 1cm 1cm 2cm}, clip,width=0.18\textwidth]{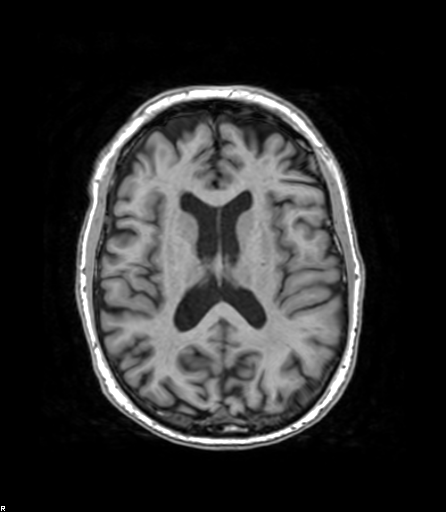}&\\
	\includegraphics[trim={0cm 0.3cm 0cm 0.3cm},clip,width=0.16\textwidth]{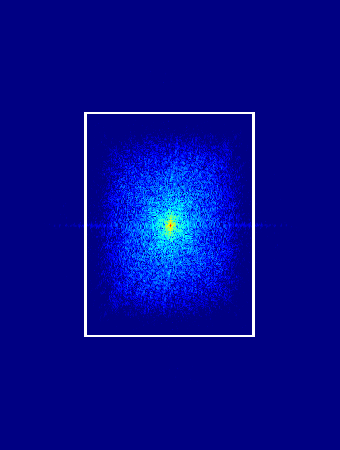}&
	\includegraphics[width=0.18\textwidth]{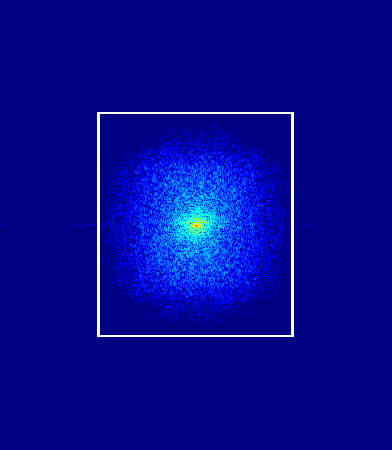}&
	\includegraphics[width=0.18\textwidth]{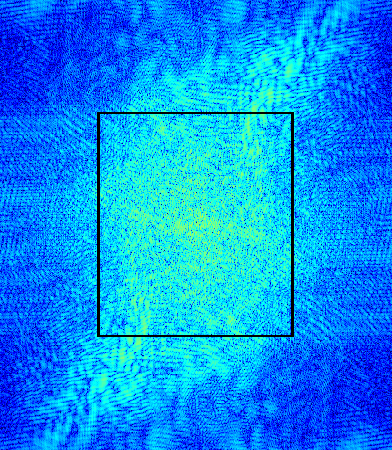}&
	\includegraphics[width=0.18\textwidth]{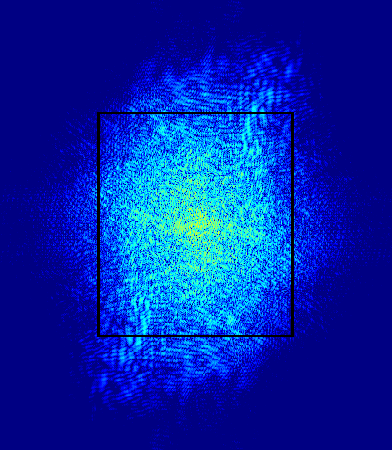}&
    \includegraphics[width=0.032\textwidth]{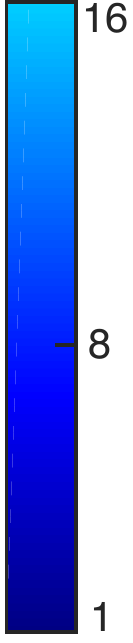}
	\end{tabular}
	\end{center}
   	\vskip -10pt
	\caption{Left to right: (Top) upsampled source $S$ and target $T$ images followed by $S$ non-linearly resampled to $T$ using sinc and Gaussian sfPSF; (Bottom) their respective frequency domain log magnitude. The white box represents the Nyquist limit before zero-padding. Note the high supra-Nyquist magnitude after sinc resampling.}
   	\vskip -10pt
	\label{imgFFT}
\end{figure}

%
\myparagraph{Frequency domain analysis under non-linear transformation} Frequency domain analysis was used to assess whether resampling images under non-linear transformation respects the Nyquist limit and to test whether the proposed method can mitigate any error. The 30 T1 MRIs were zero padded by a factor of 2 in the frequency domain (0.55$\times$0.5$\times$0.5$mm$ voxel size), doubling the representable frequencies. Fig. \ref{imgFFT} shows two upsampled images and their frequency magnitude. Note the empty power spectrum outside the Nyquist band (white box). 
29 upsampled T1 images were affinely \cite{Modat:2014jy} and then non-rigidly \cite{Modat:2010p3624} registered to the remaining image using standard algorithms. Each upsampled image was resampled to the space of the remaining image using both a three-lobed truncated sinc resampling, and the proposed Gaussian sfPSF method (see Fig. \ref{imgFFT}). The Gaussian sfPSF was set to 1.1$\times$1$\times$1$mm$ FWHM for both $\Sigma_S$ and $\Sigma_T$. Note that, as expected, supra-Nyquist frequencies are created when using sinc interpolation. These frequencies are greatly suppressed when using the proposed method. Analysis of the power spectra of the 29 resampled images showed an average power suppression of 94.4\% for frequencies above the Nyquist band when using the Gaussian PSF instead of sinc interpolation. This power suppression results in an equivalent reduction of aliasing at the original resolution. 

\begin{figure}[t!]
	\vskip -5pt
	\begin{center}
  	\tiny
    \begin{tabular}{ccc}
	\includegraphics[trim={2cm 0cm 2cm 0cm},clip,width=0.19\textwidth]{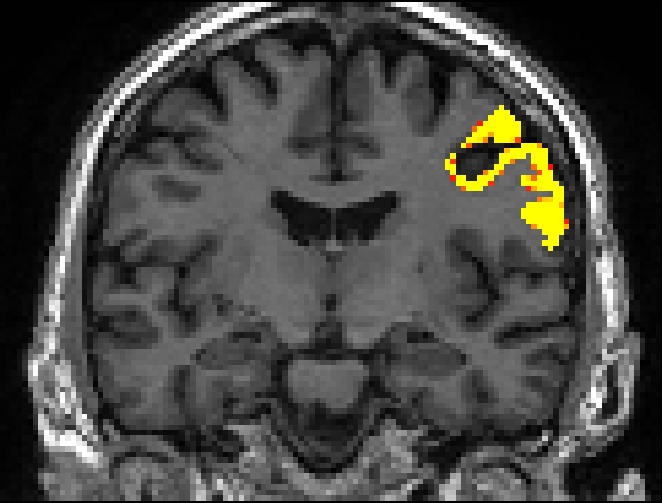}
	\includegraphics[trim={15.5cm 8cm 3.3cm 2.5cm},clip,width=0.11\textwidth]{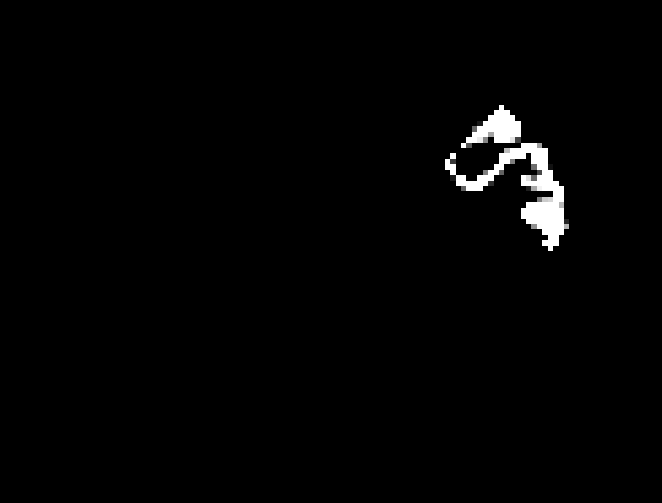}\ \ &
	\includegraphics[trim={2cm 0cm 2cm 0cm},clip,width=0.19\textwidth]{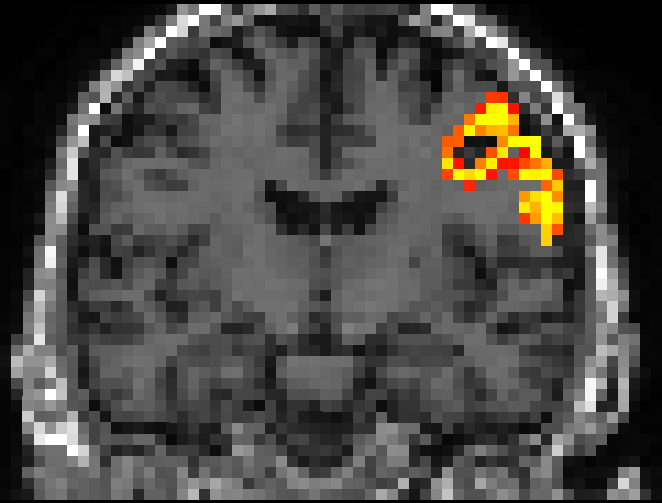}	\includegraphics[trim={15.5cm 8cm 3.3cm 2.5cm},clip,width=0.11\textwidth]{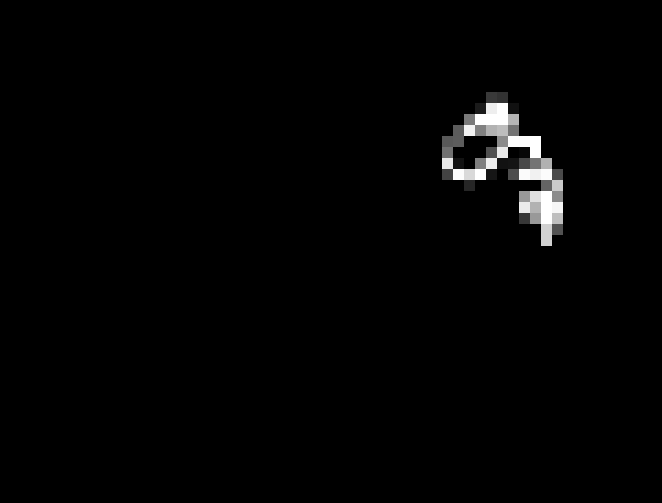}\ \ &
	\includegraphics[trim={2cm 0cm 2cm 0cm},clip,width=0.19\textwidth]{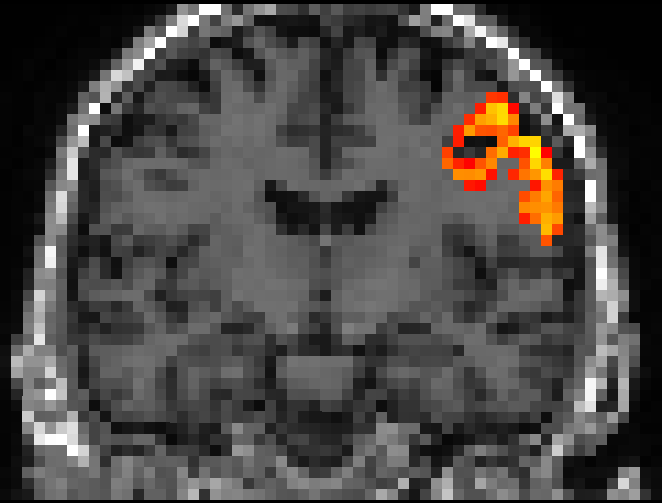}	\includegraphics[trim={15.5cm 8cm 3.3cm 2.5cm},clip,width=0.11\textwidth]{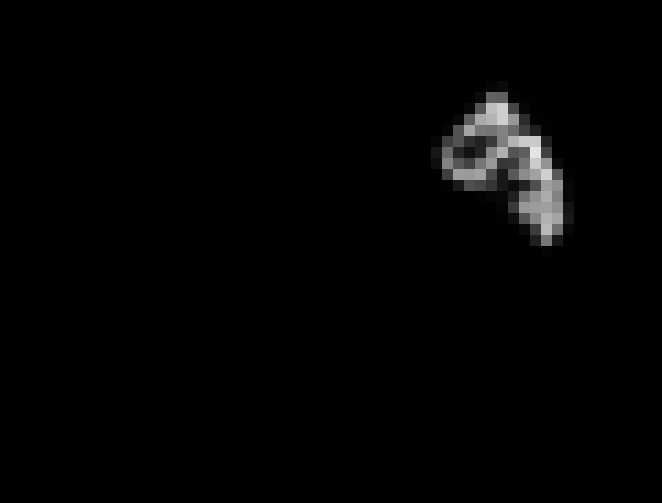}\\
        High Res Vol=14.54ml & Linear Vol=14.39ml& Gaussian sfPSF Vol=14.53ml
    \end{tabular}
	\end{center}
   	\vskip -10pt
	\caption{Left to right: A segmented region overlaid on a high resolution T1 (1.1$\times$1$\times$ 1$mm$) image (and zoomed), followed by the same segmentation resampled to the resolution of a PET image (3$\times$3$\times$3$mm$) using linear and the proposed Gaussian sfPSF. The zoomed segmentations are in gray scale between 0 and 1. Given that the cortex is commonly thinner than $3mm$, note the unrealistic large amount of voxels with segmentation probability equal to 1 at PET resolution when using linear resampling.}
    	\vskip -15pt
		\label{imgSeg}
\end{figure}

%
\myparagraph{Volume preservation and partial volume under rigid resampling} To demonstrate clinically relevant consequences of aliasing, we tested the effect of resampling within the context of segmentation propagation and partial volume estimation. Specifically, T1 images are segmented into 98 different regions using multi-atlas label propagation and fusion \cite{GIF:TMI2015} based on the Neuromorphometrics, Inc. labels. The T1 images were then rigidly registered \cite{Modat:2014jy} to the PET data, and each one of the segmented regions was resampled to the PET data using both linear interpolation and the Gaussian sfPDF. Linear was chosen here due to the signal-limited $[0,1]$ nature of probabilities. Example results are shown in Fig. \ref{imgSeg}.
\begin{figure}[b!]
	\vskip -10pt
	\begin{center}
	\includegraphics[trim=10 0 20 0,clip, width=0.46\textwidth]{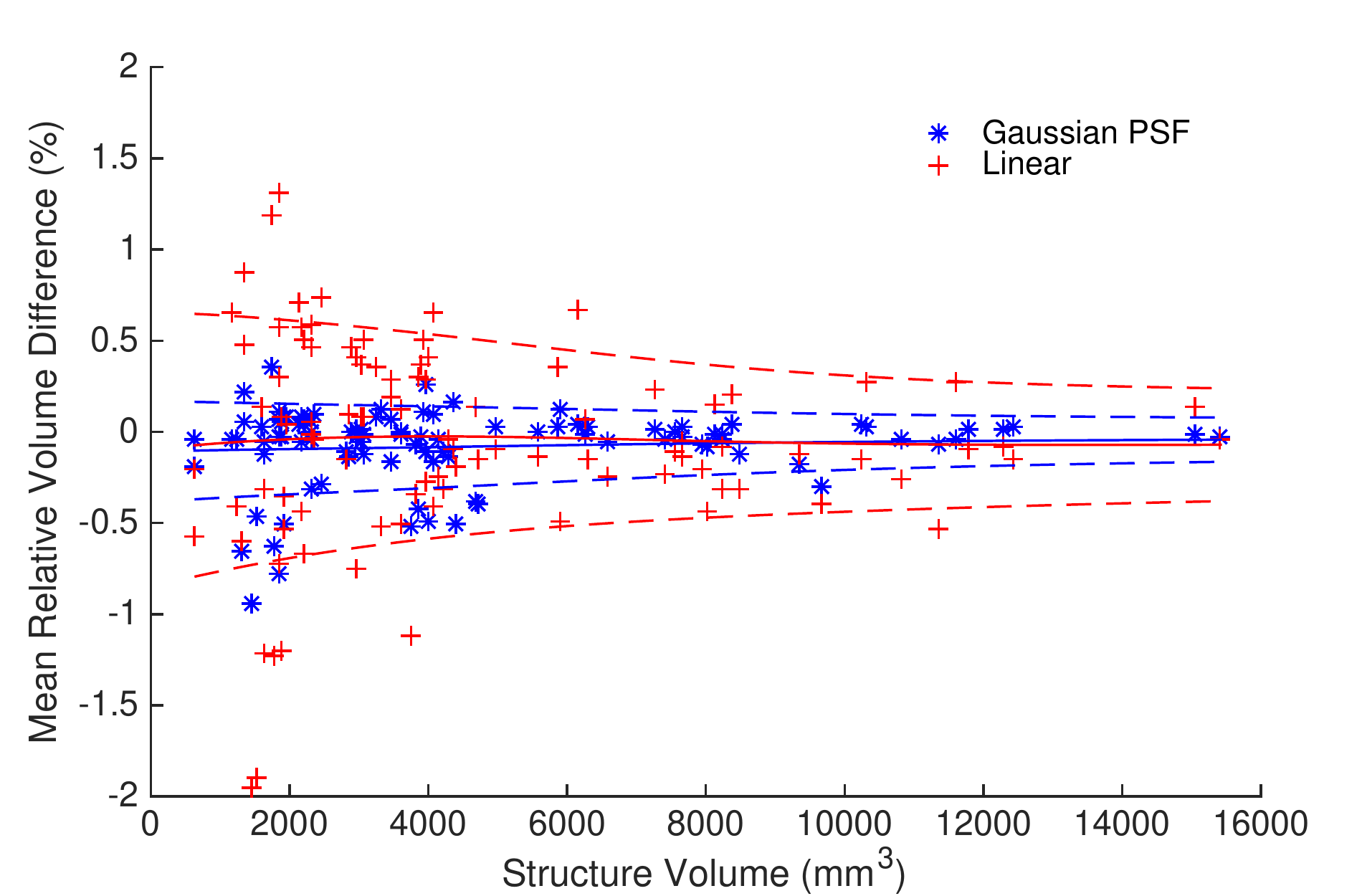}\ \ \ \ 
	\includegraphics[trim=12 0 22 0,clip,width=0.46\textwidth]{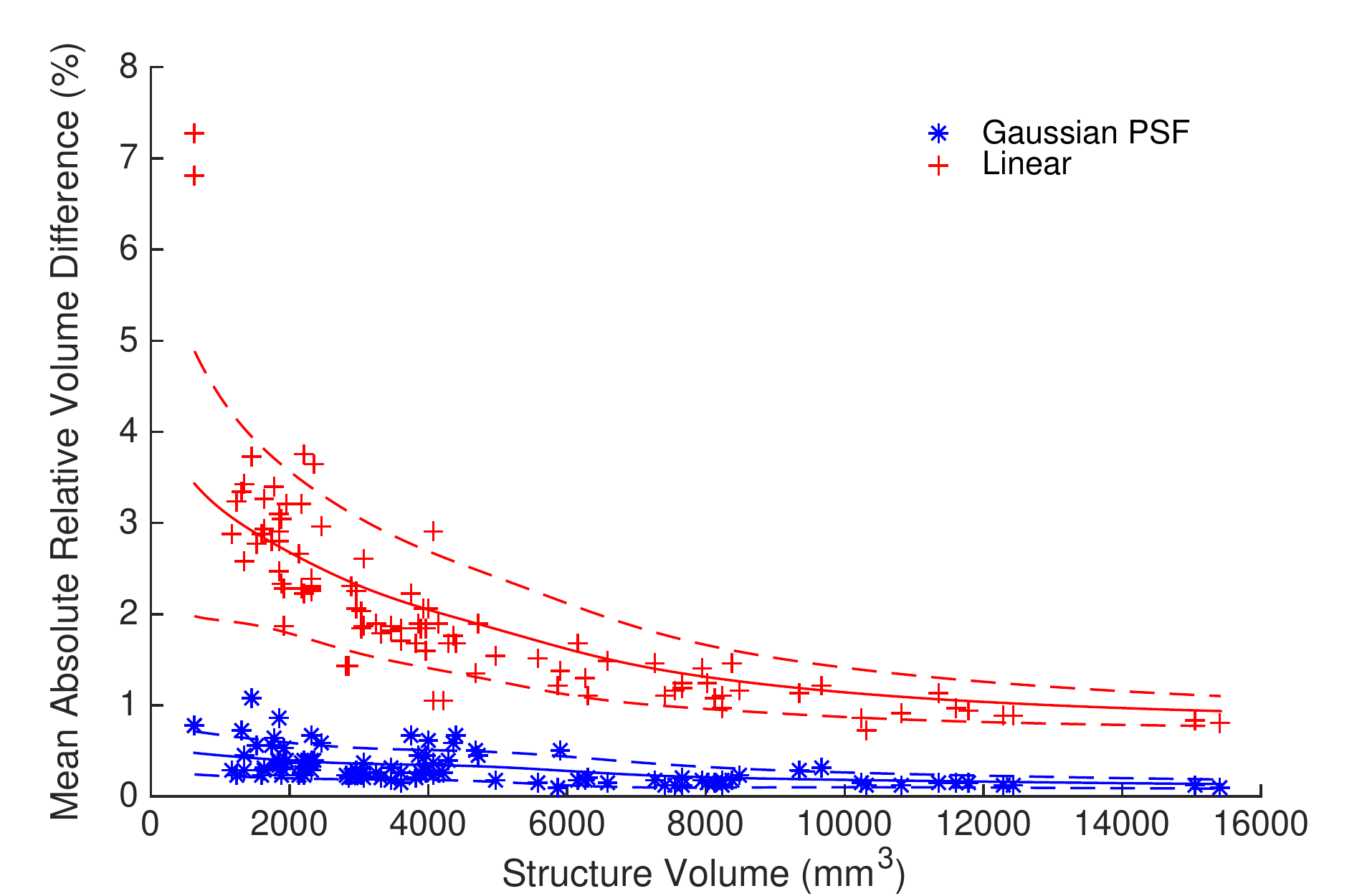}
	\end{center}
   	\vskip -12pt
	\caption{The mean RVD (Left) and mean ARVD (Right) between a low and  high resolution representation of a segmented region averaged over the population. 100 different regions are plotted against the mean volume of the region $V_{HR}$ over the population.}
    	\vskip -15pt
		\label{imgError}
\end{figure}
As all the segmentations are probabilistic, volume was estimated as the sum of all probabilities for all voxels $\mathbf{v}$ times the voxel size $\mathcal{V}$. The volume was estimated in the original T1 space ($V_{HR}$), and in the PET space after linear ($V_{TRI}$) and the Gaussian sfPSF ($V_{sfPSF}$) resampling. Similar volumes were obtained when the probabilistic segmentations were thresholded at 0.5. The relative volume difference RVD$=(V_{sfPSF}-V_{HR})/V_{HR}$ and its absolute value, ARVD $=|$RVD$|$, were estimated for each of the 30 subjects, 98 cortical regions and 2 resampling methods. The RVD and ARVD were averaged over all subjects for each region, resulting in 98 mean RVD and ARVD values per region, per method. These values are plotted in Fig. \ref{imgError}. A Parzen window mean and STD are also plotted in the same figure. Both resampling methods are unbiased according to the mean RVD. However, the Gaussian sfPSF method provides significantly lower ($p<10^{-4}$ - Wilcoxon signed-rank test) errors in terms of mean ARVD for all regions, especially in smaller regions where linear resampling can introduce up to 8\% mean absolute volume difference. An 8\% mean difference implies even larger local errors, resulting in detrimental effects in partial volume/ compartment modelling, and when estimating biomarkers that rely on structural segmentations.

%

\section{Conclusion}
The presented work explores aliasing in medical image resampling and proposes a new interpolation technique that matches the scale factor PSF given arbitrary grid sizes and non-linear transformations. We demonstrate the advantages of using Gaussian sfPSF resampling, both in terms of Nyquist limit and volume preservation, when compared with common resampling techniques. Future work will involve deploying the proposed methodology within image registration algorithms and verifying the impact of the sfPSF in partial volume and compartment modelling. An implementation will be made available at the time of publication. 

\myparagraph{Acknowledgements}
This work was supported by the EPSRC (EP/H046410/1, EP/J020990/1, EP/K005278), the MRC (MR/J01107X/1), the EU-FP7 (FP7-ICT-2011-9-601055), the NIHR BRC UCLH/UCL HII (BW.mn.BRC10269), and the UCL Leonard Wolfson Experimental Neurology Centre (PR/ylr/18575). 

%
\bibliographystyle{splncs}
\bibliography{JNAbrv,PV,PV2,biblio}

\clearpage
\section*{Draft Appendix}

Let $S$ and $T$ be 2 symmetric positive definite (SPD) matrices representing covariance matrices. We are looking for the \emph{smallest} symmetric positive semi-definite (SPSD) matrix $P$ such that $S+P$ is larger than $T$.

%
\subsection*{Some useful tools and properties}
A matrix $X$ is SPSD if
\begin{equation}
\forall v, v^\intercal \cdot X \cdot v \geq 0
\end{equation}
We then write
\begin{equation}
X \succeq 0
\end{equation}
Furthermore, we write $X \succeq Y$ iff $X-Y \succeq 0$.

A matrix $X$ is SPD if
\begin{equation}
\forall v \neq 0, v^\intercal \cdot X \cdot v > 0
\end{equation}
We then write
\begin{equation}
X \succ 0
\end{equation}
We also write \[X \succ Y \textrm{~iff~} X-Y \succ 0\].

If $Z$ is invertible, then $Z^\intercal \cdot X \cdot Z \succeq 0$ (resp. $\succ 0$) iff $X \succeq 0$ (resp. $\succ 0$).
The spectral decomposition of a SPSD matrix is given by
\begin{equation}
X = Z(X) \cdot \Lambda(X) \cdot Z(X)^\intercal
\end{equation}
where $Z(X) \cdot Z(X)^\intercal = \Id$
and $\Lambda(X) = \diag(\lambda_i(X))$. Note that for any matrix $Y$,
\begin{equation}
 \det( Z(X) \cdot Y \cdot Z(X)^\intercal ) = \det(Y)
\end{equation}

%
\subsection*{Smallest SPSD matrix increment}

\subsubsection*{A matrix norm minimization view.}
We are looking for an SPSD matrix $P$ such that $S+P$ is close to $T$. In other words, we want $P$ to be close to $T-S$. We rephrase this more formally as minimizing the Frobenius norm $\norm{(S+P)-T}=\norm{P-(T-S)}$ subject to the condition $P \succeq 0$.

Following \cite{Higham:LAA:1988}, the solution to this problem is given by:
\begin{equation}\label{eq:Pnorm}
P = Z(T-S) \cdot \max( \Lambda(T-S), 0) \cdot Z(T-S)^\intercal
\end{equation}
While this provides a nice closed-form solution. It is unclear whether the Frobenius norm really captures the discrepancy in term of covariance matrix.

This solution still has interesting properties. We indeed have $S+P \succeq S$ and $S+P \succeq T$. Indeed, from \eqref{eq:Pnorm}, we easily see that $P - (T-S) \succeq 0$

\subsubsection*{A geometric view.}
We are looking for an SPSD matrix $P$ that minimizes the volume of $S+P$ (i.e. $\det(S+P)$) under the constraints that $P \succeq 0$, $S+P \succeq S$ (which is trivially obtained from $P \succeq 0$) and $S+P \succeq T$.

Let us first generalize the problem. We now simply look for an SPD matrix $Q$ that minimizes $\det(Q)$ subject to the conditions $Q \succeq S$ and $Q \succeq T$.
Let us consider the spectral decomposition of $S$ and use the corresponding spectral coordinate system:
\begin{equation}
 X' = Z(S)^\intercal \cdot X \cdot Z(S)
\end{equation}
Note that we have $S'= \Lambda(S)$.
Our problem is equivalent to minimizing $\det(Q')=\det(Q)$ subject to the conditions $Q' \succeq S'=\Lambda(S)$ and $Q' \succeq T'$.
\begin{figure}[h!tb]
  \begin{center}
  \includegraphics[width=0.4\linewidth]{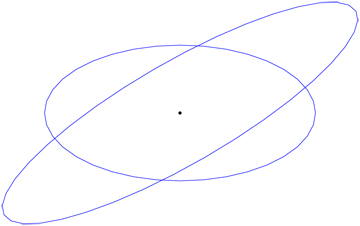}
  \caption{Covariances in coordinate system of $S$.}
  \label{im:1}
  \end{center}
\end{figure}

Let us now rescale the space with a volume-preserving transformation to turn $\Lambda(S)$ into a scalar matrix. Let 
\begin{equation}
\Lambda_N(S) = \det(S)^{1/(2*Dim)}\cdot\diag(\lambda_i^{-1/2}(S))
\end{equation}
where $Dim$ is the dimension of the space and $\det(\Lambda_N(S)) = 1$. Let
\begin{equation}
 X'' = \Lambda_N(S) \cdot X' \cdot \Lambda_N(S) = \Lambda_N(S) \cdot Z(X)^\intercal \cdot X \cdot Z(X) \cdot \Lambda_N(S)
\end{equation}
Note that we have $S''=\det(S)^{1/Dim}\Id$.
Our problem become equivalent to that of minimizing $\det(Q'')=\det(Q)$ subject to the constraints $Q'' \succeq S''=\det(S)^{1/Dim}\Id$ and $Q'' \succeq T''$.
\begin{figure}[h!tb]
  \begin{center}
  \includegraphics[width=0.4\linewidth]{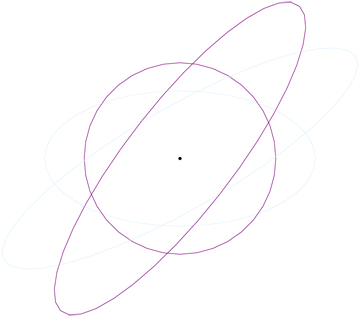}
  \caption{Volume-preserving rescaling to make $S$ isotropic.}
  \label{im:2}
  \end{center}
\end{figure}

Let us now consider the spectral decomposition of $T''$ and use the corresponding spectral coordinate system:
\begin{equation}
 X''' = Z(T'')^\intercal \cdot X'' \cdot Z(T'')
\end{equation}
Note that we have $T''' = \Lambda(T'')$.
Our problem become equivalent to that of minimizing $\det(Q''')=\det(Q)$ subject to the constraints $Q''' \succeq S'''=\det(S)^{1/Dim}\Id$ and $Q''' \succeq T'''= \Lambda(T'')$.
Note that both $\det(S)^{1/Dim}\Id$ and $\Lambda(T'')$ are axis-aligned, \ie diagonal.
\begin{figure}[h!tb]
  \begin{center}
  \includegraphics[width=0.4\linewidth]{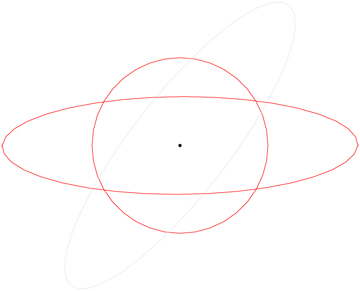}
  \caption{Rotation to make both covariances axis-aligned.}
  \label{im:4}
  \end{center}
\end{figure}

For symmetry-reasons, and because of the uniqueness of the solution \cite{Schrocker:CAGD:2008}, it has to also be axis-aligned, \ie diagonal: $Q'''=\diag(\lambda_i(Q'''))$. We find that our problem is now equivalent to minimizing $\det(Q''')=\prod \lambda_i(Q''')$ subject to the conditions $\lambda_i(Q''')\geq\det(S)^{1/Dim}$ and $\lambda_i(Q''')\geq\lambda_i(T'')$.
The solution to this problem is obtained by taking:
\begin{equation}
\lambda_i(Q''') = \max(\det(S)^{1/Dim},\lambda_i(T''))
\end{equation}
\begin{figure}[h!tb]
  \begin{center}
  \includegraphics[width=0.4\linewidth]{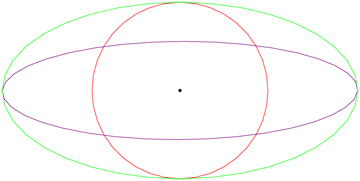}
  \caption{Best covariance in this space?}
  \label{im:5}
  \end{center}
\end{figure}

Now we need to roll back the transformations and get
\begin{multline}\label{eq:Qgeom}
Q = Z(S) \cdot \Lambda_N^{-1}(S) \cdot Z(T'') \\
\cdot\max(\det(S)^{1/Dim}\Id,\Lambda(T'')) \cdot Z(T'')^\intercal \cdot \Lambda_N^{-1}(S) \cdot Z(S)^\intercal
\end{multline}

Equivalently, focusing on $T$ first, we get:
\begin{multline}\label{eq:QgeomT}
Q = Z(T) \cdot \Lambda_N^{-1}(T) \cdot Z(S'') \\
\cdot\max(\det(T)^{1/Dim}\Id,\Lambda(S'')) \cdot Z(S'')^\intercal \cdot \Lambda_N^{-1}(T) \cdot Z(T)^\intercal
\end{multline}

\paragraph{Axis-aligned \texorpdfstring{$T$}{T}.}
If $T$ is axis aligned, \eqref{eq:QgeomT} simplifies to:
\begin{align}\label{eq:QgeomAxisT}
Q = T^{1/2} \cdot Z(S''')\cdot\max(\Id,\Lambda(S''')) \cdot Z(S''')^\intercal \cdot T^{1/2}
\end{align}
with
\begin{align}
S''' = T^{-1/2} \cdot S \cdot T^{-1/2}
\end{align}

\subsubsection*{Relating the matrix norm and the geometric views.}
\begin{figure}[h!tb]
  \begin{center}
  \includegraphics[width=0.8\linewidth]{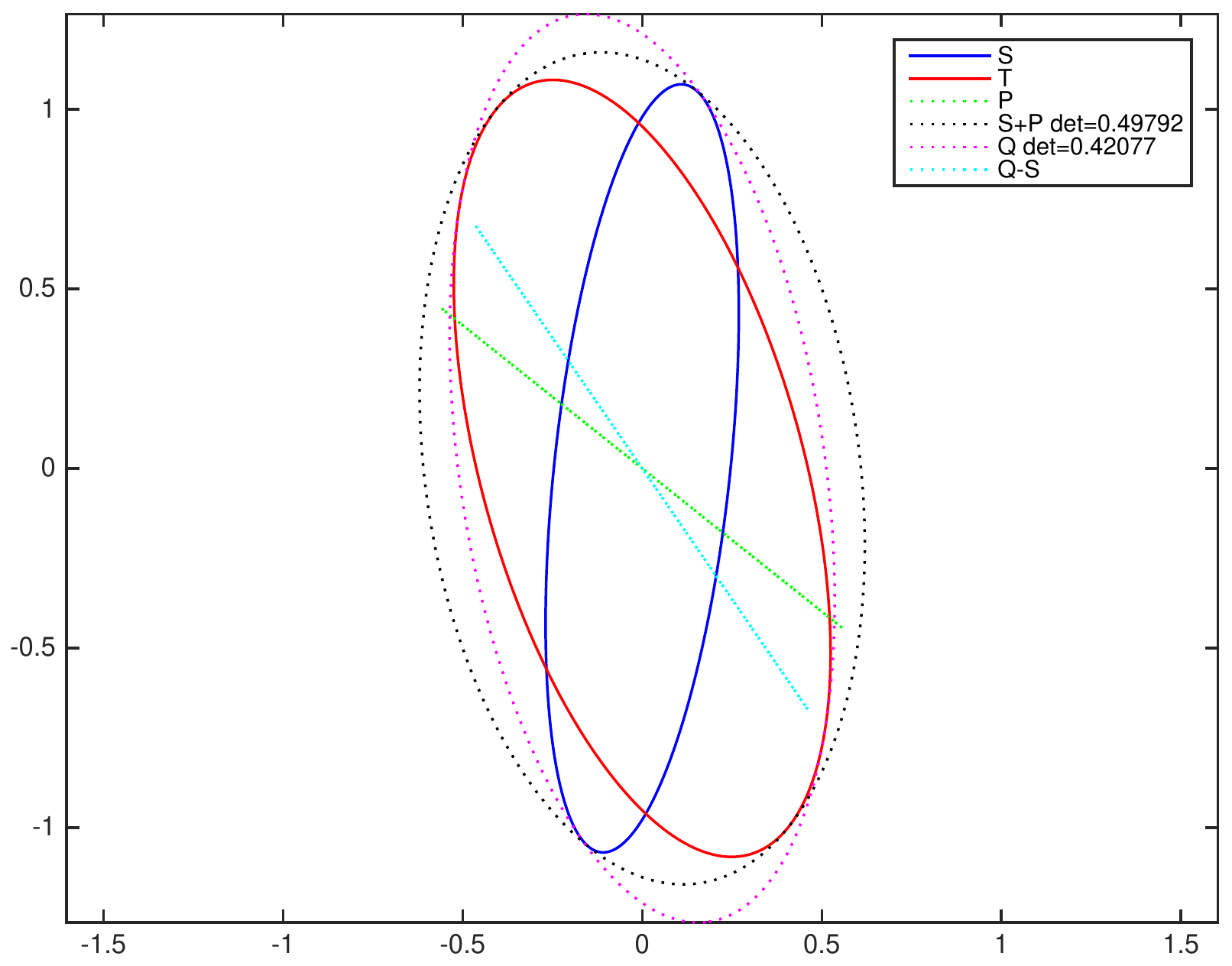}
  \caption{Summary of all ellipses}
  \label{im:all}
  \end{center}
\end{figure}

At this stage it's unclear how \eqref{eq:Pnorm} and \eqref{eq:Qgeom} relate to each other. Numerical simulations shows a difference but the results are quite close to each other. \Figref{im:all} shows an example with a rather large difference. The geometric method indeed leads to smaller volume but this comes at the cost of increasing the anisotropy. The matrix norm method produces more intuitive results.
Some interesting references include \cite{Higham:LAA:1988,Schrocker:CAGD:2008}
and
\url{http://mathoverflow.net/questions/120925}.

\end{document}